\documentclass[5p,twocolumn,times,number,pdflatex]{elsarticle}

\usepackage{hyperref}
\usepackage{amsmath,amssymb}
\usepackage{xcolor,overpic}

\journal{Elsevier}

\def\pmbanner{\vskip39pt{}\vskip39pt{}\vskip20pt}

\begin{document}

\begin{frontmatter}

\title{\pmbanner The detectors of the SHiP experiment at CERN}

\author[a1]{E.~Graverini}
\ead{elena.graverini@cern.ch}
\author{ on behalf of the SHiP collaboration}
\address[a1]{Universit\"at Z\"urich}




\begin{abstract}
SHiP is a proposed general purpose fixed target facility at the CERN SPS accelerator.
The main focus will be the physics of the Hidden Sector, \textit{i.e.} search for heavy neutrinos, dark photons and other long lived very weakly interacting particles.
A dedicated detector, based on a long vacuum tank followed by a spectrometer and particle identification detectors, will allow probing a variety of models with exotic particles in the GeV mass range.
Another dedicated detector will allow the study of Standard Model neutrino cross-sections and angular distribution, and allow detection of light dark matter with world leading sensitivity.
\end{abstract}


\end{frontmatter}

\section{The SHiP experiment}
Search for Hidden Particles (SHiP) is a proposed general purpose beam dump facility, aimed at searching for very weakly interacting long-lived particles, collectively referred to as the Hidden Sector (HS).
An integrated flux of $2\times 10^{20}$ protons colliding onto a heavy target will allow SHiP to probe a great variety of New Physics models, improving the current limits by several orders of magnitude in 5 years of nominal operation~\cite{TP,PP}.
Heavy Neutral Leptons
(HNLs), right-handed partners of the Standard
Model (SM) neutrinos, will be searched for in decays of beauty and charm mesons. The existence of such particles is
strongly motivated by theory, as they can simultaneously
solve multiple problems left open by the SM.
In the Neutrino Minimal Standard Model ($\nu$MSM), HNLs
explain the baryon asymmetry of the Universe, account
for the pattern of neutrino masses and oscillations and
provide a dark matter candidate~\cite{Asaka:2005an}.
A redundant system of background tagging detectors
will make SHiP a zero-background experiment.
In addition, the peculiar layout of the SHiP facility also make it a very efficient Standard Model neutrino factory. Therefore,
a dedicated detector nicknamed $i$SHiP will be installed upstream of the decay volume for hidden particles, aimed at studying the $\nu_\tau$ properties.
The SHiP Technical Proposal~\cite{TP} and physics case~\cite{PP} have been positively reviewed by the CERN SPS Committee in 2015, and the collaboration has been prompted to produce a Comprehensive Design Report.

\subsection{The SHiP detector}
\begin{figure}
	\centering\includegraphics[width=\linewidth]{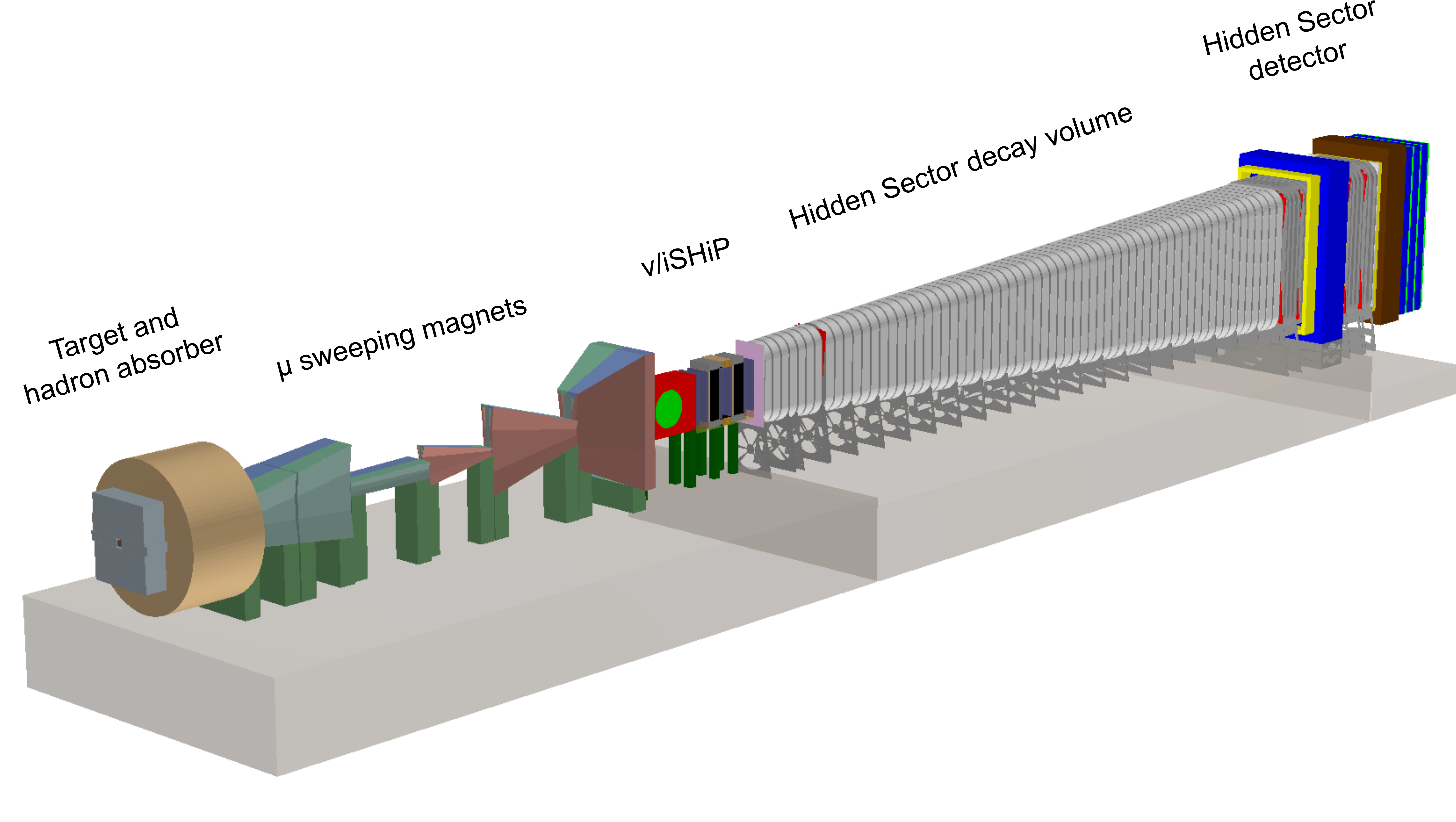}
	\caption{\textsc{Geant4}~\cite{geant4} implementation of the proposed SHiP layout.}\label{fig:layout}
\end{figure}

A scheme of the SHiP facility is illustrated in \figurename~\ref{fig:layout}.
A dedicated beam line, extracted from the SPS using the same transfer line as the other CERN North Area experiments, will convey a 400 GeV$/c$ proton beam.
The beam will be stopped in a heavy Molybdenum-Tungsten target, which is followed by a hadron stopper and by a system of shielding magnets sweeping residual muons away from the detector fiducial area.

The target of the $i$SHiP detector will be placed in the muon free area, and it will be composed of bricks of laminated lead and emulsions, interleaved with a scintillating fiber spectrometer and followed by a muon spectrometer, allowing to measure the momentum and identify the flavour of scattering leptons.

The main element of Hidden Sector detector will be a $\sim$50~m long decay volume, kept at a pressure of $10^{-6}$~atm to minimize the probability of $\nu$ interactions in air. 
The decay volume is contained in a pyramidal frustum shaped vessel, with maximum transversal dimensions of $5\times 10$~m$^2$.
The length and shape of the vessel have been optimised by maximizing the acceptance to the hidden particle decay products given the dimensions of the muon-free area achieved by the shielding magnet, which shape and power were in turn optimized based on expected cost and performance.
The vessel walls enclose segmented 30~cm thick sections of liquid scintillator aimed at tagging particles coming from outside.
An upstream straw detector, placed in vacuum at a distance of 5~m from the vessel entry lid, will identify events initiated in the material upstream of the fiducial decay volume.

The vacuum vessel will be followed by a detector for the charged hidden particle daughters.
Momentum information will be provided by a tracking system placed in vacuum at the end of the decay volume, made of four stations of 5~m long straw tubes placed on either side of a 1~Tm magnet.
A high-accuracy timing detector will be installed, aimed at reducing the yield of combinatorial events. Two technologies are being considered: plastic scintillating bars and multigap resistive plate
chambers (MRPC). Both technologies can be based on
existing and well-studied designs and can reach a
time resolution of 100~ps~\cite{TP}.
Particle identification will be provided by Shashlik electromagnetic and hadronic calorimeters, followed by a muon system made of four active layers interlaced with iron.

\subsection{Background suppression}

Due to the small coupling expected between hidden particles and the SM, production rates of $10^{-10}$ or lower are expected.
The vast majority of SM particles produced in the target are stopped by the hadron absorber or swept out of acceptance by the muon shield; however, residual muons or other particles can still enter the decay volume and produce a signal in the HS detector. These events produce however a signal in the liquid scintillator tagger, too, and are therefore removed.
The main background to the hidden particle signal originates instead from the scattering of SM particles (especially neutrinos) in the vicinity of the HS decay volume, producing long-lived neutral mesons such as $K_S$. These particles nevertheless decay before the upstream straw tagger or have a vertex very close to the vessel walls, with the reconstructed candidate not pointing to the beam target. 
Random combination of tracks can also mimic HS events, but the use of a high resolution timing detector will reduce this background to a negligible level.
Thorough studies reported in~\cite{TP} have found no evidence of any irreducible background, allowing to set an upper limit of 0.1 expected background events during the whole 5 years SHiP run.

\subsection{Physics performance}
\begin{figure}[h]
	\centering
	\begin{overpic}[width=.7\linewidth]{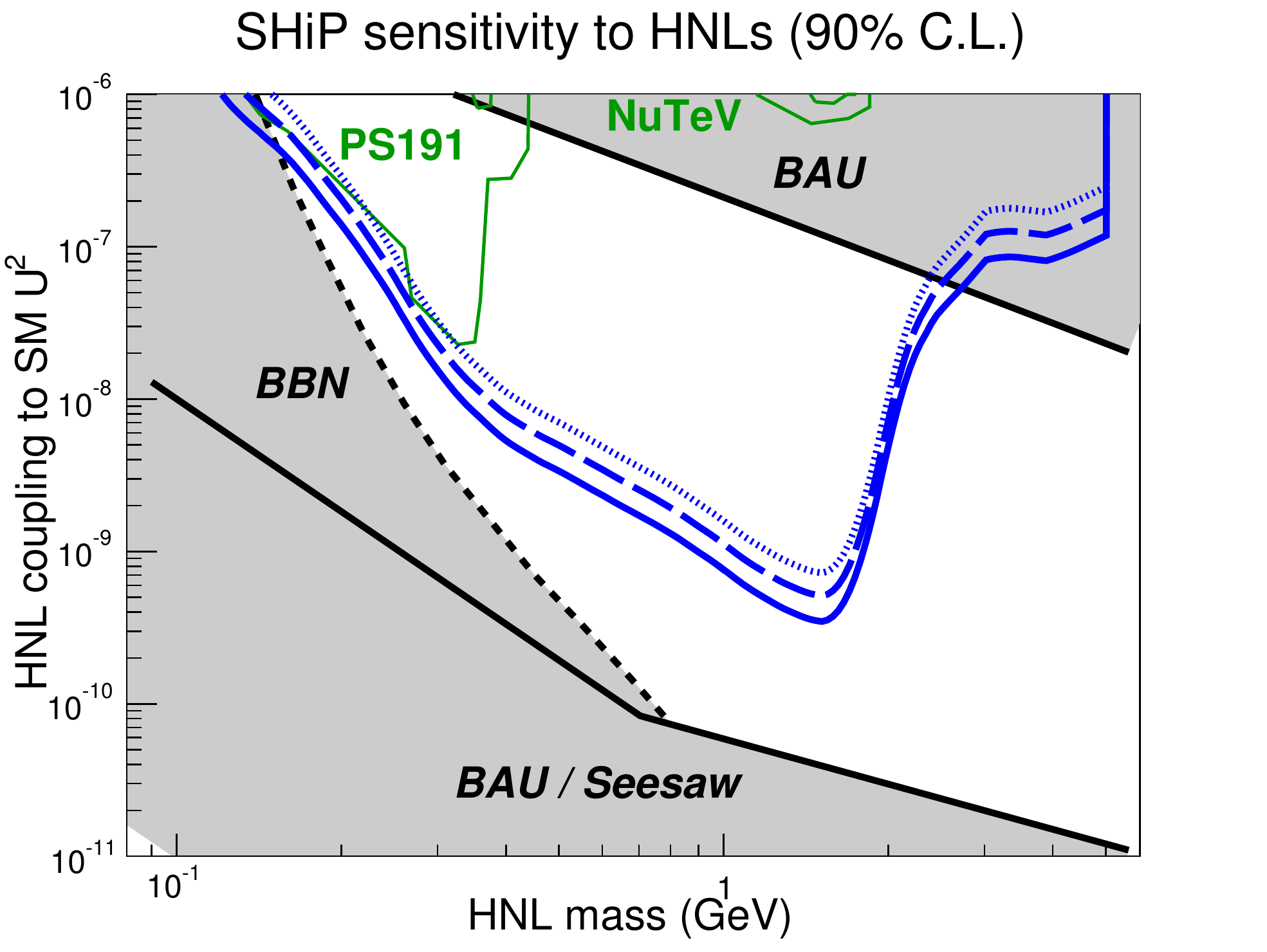}\put(55, 40){\textcolor{blue}{\it SHiP}}\end{overpic}
    \caption{SHiP's discovery potential in the parameter space of the $\nu$MSM~\cite{TP}.}\label{fig:sens}
    \vspace*{2mm}
    \begin{overpic}[width=.8\linewidth]{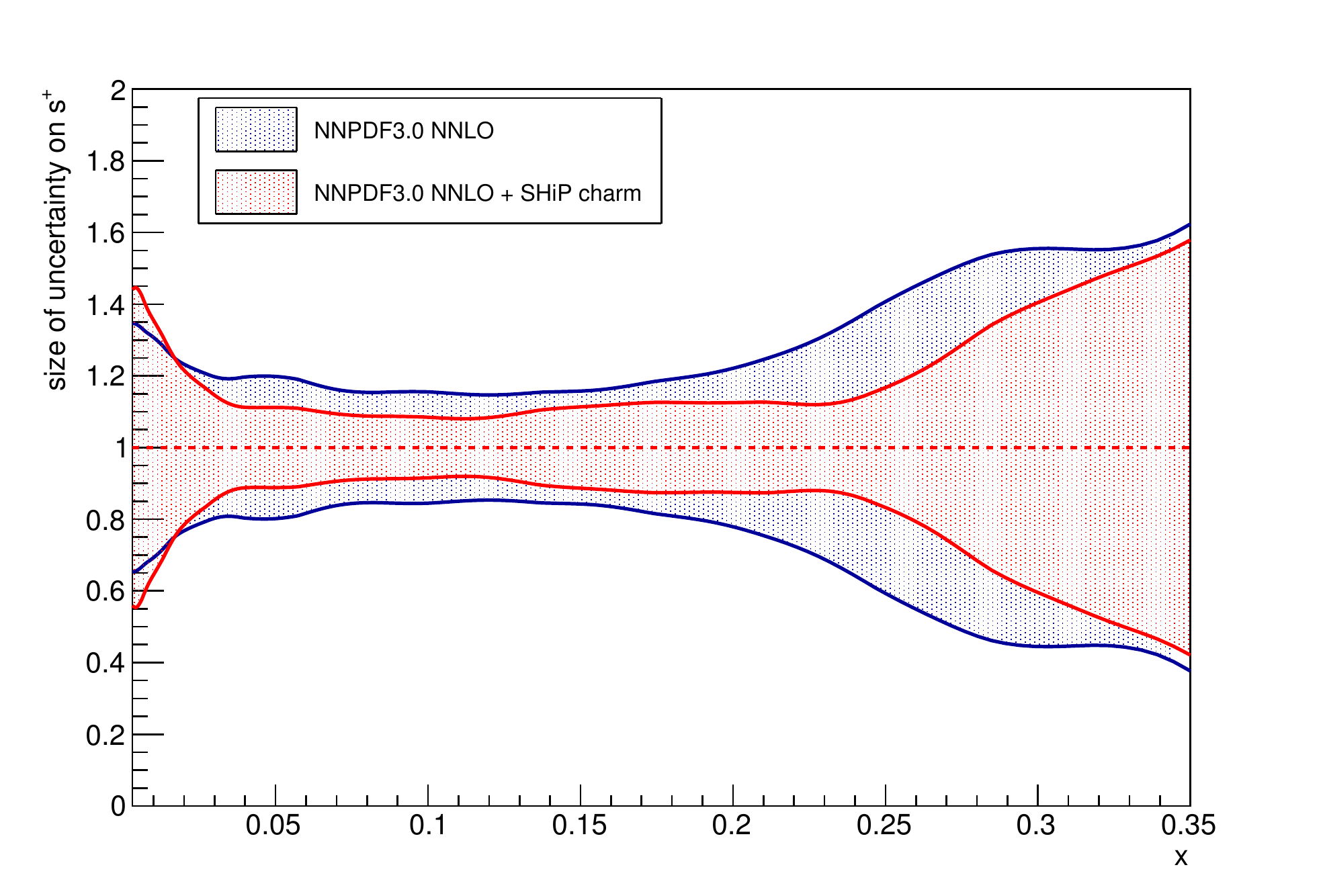}\end{overpic} \vspace*{-2.mm}
	\caption{SHiP contribution to the strange quark sea uncertainty bands~\cite{PP}.}\label{fig:nu}
\end{figure}
With this level of background, the sensitivity of the SHiP detector is illustrated in \figurename~\ref{fig:sens} for the $\nu$MSM, as a benchmark New Physics model with light long-lived particles. The expected sensitivities to hidden particles exceed current limits by several orders of magnitude, depending on the model~\cite{PP}.

The high intensity proton beam dump will produce a large flux of neutrino of all three flavours impinging on the $i$SHiP target. The $\bar{\nu}_\tau$ will be observed for the first time, and the cross sections of $\nu_\tau$ and $\bar{\nu}_\tau$ will be measured with high precision.  Such a neutrino flux will allow to measure for the first time the $F_4$ and $F_5$ structure functions governing charged current $\nu_\tau$ interactions. In addition, neutrino-nucleon deep inelastic scattering events can improve our understanding of the flavour composition of the nucleon, allowing to study the strange quark content. About 2 millions and 1 million events are expected from $\nu_\mu$ and $\nu_e$ scattering, respectively, which would significantly improve our knowledge of the strange sea as shown in \figurename~\ref{fig:nu} and as documented in~\cite{PP}.
The $i$SHiP detector will also be ideally suited to detect the signal generated by light dark matter particles scattering on the electrons of the emulsion target. The background from neutrino scattering can be reduced to less than 300 expected events during the whole SHiP run. SHiP is expected to be sensitive in portions of the light dark matter parameter space not accessible to previous experiments~\cite{PP}.

\bibliography{mybibfile}

\end{document}